\documentclass[prl,aps,superscriptaddress,showpacs,showkeys,twocolumn,letterpaper]{revtex4-1}
\usepackage{amsmath}
\usepackage[breaklinks]{hyperref}
\usepackage {graphicx}
\usepackage{multirow}
\usepackage{times}
\newcommand {\nc} {\newcommand}
\newcommand {\rn} {\renewcommand}

\nc{\bittot}{Bi$_2$Sr$_2$CaCu$_2$O$_{8+\delta}$}
\nc{\bitset}{Bi$_2$Se$_3$}
\nc{\capa}{\mathcal{C}_n}
\nc{\caparison}{\mathcal{C}_n(\vec{k}, \omega)}
\nc{\cuotwo}{CuO$_2$}
\nc{\ef}{E_F}
\nc{\ek}{\varepsilon(\kvec)}
\nc{\ep}{E_p}
\nc{\gkw}{G \kw}
\nc{\hightc}{high T$_c$}
\nc{\ikw}{I \kw}
\rn{\Im}{\mathrm{Im}\,}
\nc{\ims}{\Im \Sigma}
\nc{\kf}{k_F}
\nc{\kvec}{\vec{k}}
\nc{\skw}{\Sigma \kw}
\nc{\res}{\mathrm {Re}\Sigma}
\nc{\lsco}{La$_{2-x}$Sr$_x$CuO$_4$}
\nc{\scco}{Sr$_{14-x}$Ca$_{x}$Cu$_{24}$O$_{41}$}
\nc{\sco}{SrCu$_2$O$_3$}
\nc{\ladderlayer}{Cu$_2$O$_3$}
\nc{\chainlayer}{CuO$_2$}
\nc{\tc}{T$_c$}
\nc{\tj}{$t$-$J$}
\nc{\tll}{\scco}
\nc{\tbd}{??}
\nc{\tone}{$t_1$}
\nc{\ttwo}{$t_2$}
\nc{\toneperp}{\ensuremath{t_{1,\perp}}}
\nc{\ttwoperp}{\ensuremath{t_{2,\perp}}}
\nc{\fr}{\frac}
\nc{\lf}{\left}
\nc{\rg}{\right}
\nc{\xc}{C$_3$--X--C}
\nc{\rr}{R$_2$--R}

\begin{document}

\title {A spectroscopic fingerprint of electron correlation in high temperature superconductors}
\author {G.-H. Gweon}
\email {Corresponding author, gweon@ucsc.edu.}
\affiliation {Department of Physics, University of California, Santa Cruz, CA 95064}

\author {G.-D. Gu}
\author {J. Schneeloch}
\author {R. D. Zhong}
\affiliation {Brookhaven National Laboratory, Upton, NY 11973}
\author {T. S. Liu}
\affiliation {Brookhaven National Laboratory, Upton, NY 11973}
\affiliation {School of Chemical Engineering and Environment, North University of China}

\date {\today}
\begin {abstract} 
The so-called ``strange metal phase'' \cite{anderson_strange_2006} of high temperature (\hightc{}) superconductors remains at the heart of the \hightc{} mystery.  Better experimental data and insightful theoretical work would improve our understanding of this enigmatic phase.  In particular, the recent advance in angle resolved photoelectron spectroscopy (ARPES) \cite{koralek_laser_2006,liu_development_2008}, incorporating low photon energies ($\approx 7$ eV), has given a much more refined view of the many body interaction in these materials.  Here, we report a new ARPES feature of \bittot{} that we demonstrate to have the key ability to distinguish between different classes of theories of the normal state.  This feature---the anomaly in the nodal many body density of states (nMBDOS)---is clearly observed in the low energy ARPES data, but also observed in more conventional high energy ARPES data, when a sufficient temperature range is covered.  We show that key characteristics of this anomaly are explained by a strong electron correlation model; the electron-hole asymmetry and the momentum dependent self energy emerge as key required ingredients.
\end {abstract} 


\maketitle

The strange metal phase \cite{anderson_strange_2006} of high temperature superconductors is characterized by various anomalous transport and spectroscopic characteristics \cite{anderson_strange_2006,varma_phenomenology_1989}, still to be understood clearly.  Among these experimental techniques, angle resolved photoelectron spectroscopy (ARPES) provides special insight.

Recent advances in theory \cite{shastry_extremely_2011} and experiment \cite{gweon_extremely_2011,matsuyama_phenomenological_2013} have renewed interest in this topic.  These recent works \cite{gweon_extremely_2011,matsuyama_phenomenological_2013} adopted the line shape fit approach, as in previous works \cite{kaminski_momentum_2005,casey_accurate_2008}, and showed that the ARPES data can be understood very comprehensively using the new extremely correlated Fermi liquid (ECFL) \cite{shastry_extremely_2011,shastry_extremely_2012} framework.  Taking a more global view, one can ask whether it is possible to compare different line shape fit analyses \cite{kaminski_momentum_2005,casey_accurate_2008,gweon_extremely_2011,matsuyama_phenomenological_2013}.  While such work may be possible, it is a complex time-consuming process.  More fundamentally, a bigger issue is that it is not always clear how valid certain assumptions made in each many body theory adopted for line shape description really are.  Therefore, finding a robust experimental finger print feature that can discern between different theoretical models in regards to their key assumptions would be very valuable.  Finding such a feature would help determine whether the strange normal phase is a true non-Fermi liquid state \cite{varma_phenomenology_1989,anderson_theory_1997,anderson_strange_2006} or a very unconventional Fermi liquid state \cite{shastry_extremely_2011}.


Here, we report such a feature unearthed from a large set of ARPES data, taken as a function of wide ranges of photon energy, doping, and temperature.  While the feature itself is simple---it is the angle integrated nodal cut data, corresponding to the nodal many-body density of states (nMBDOS)---its anomalous characters have not been recognized prior to this work, to our knowledge.  As we shall see, our previous works \cite{gweon_extremely_2011,matsuyama_phenomenological_2013} happen to connect nicely to this work, whose new findings were made possible primarily by the rare focus on high temperatures.

\begin{figure*}[t] 
\includegraphics[width=0.97\textwidth]{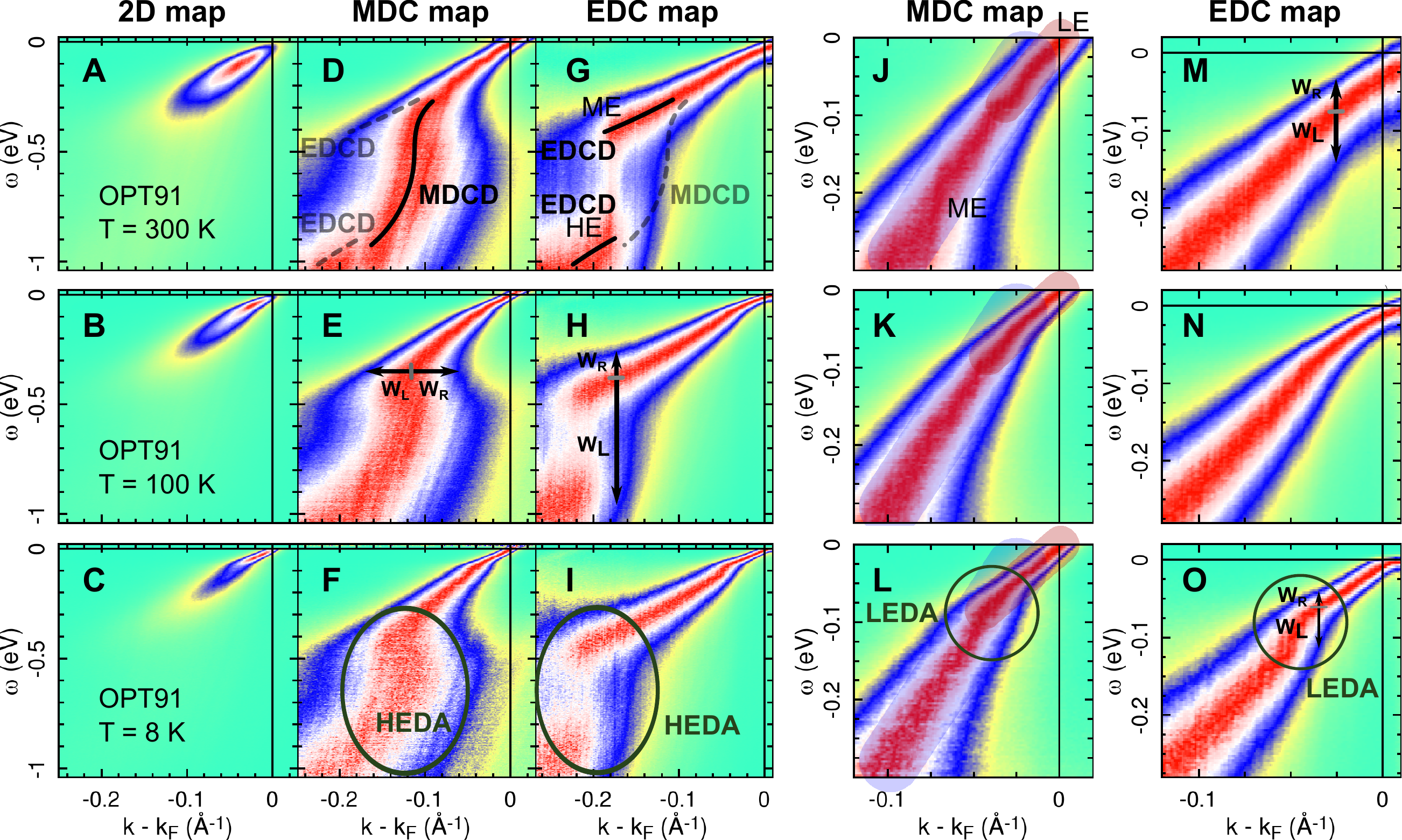}
\caption{{\bf Characteristics of the low energy ARPES data.}
ARPES data taken on an optimally doped \bittot{} sample ($T_c = 91$ K; OPT91) at three sample temperatures ($T$), using photons of energy 7.5 eV\@.  The top row corresponds to $T = 300$ K, the middle row to 100 K, and the bottom row to 8 K\@.  All panels in a given row correspond to the same ARPES data, $I(k,\omega, T)$: they differ only in terms of the format in which the color scale is used or what ranges of $k,\omega$ are used.\\
\hspace*{2em} The color scale is such that the intensity decreases from red (maximum), to blue (half maximum), and to green (minimum).  The three different formats used here---2D map, MDC map, and EDC map---are distinguished by how this color scale is applied.  In ``2D maps'' (first column), this color scale is applied to $I(k,\omega, T)$, globally in the full two dimensional domain defined by $k,\omega$ axes: this is the usual format in which ARPES data are represented as a map.  We use two additional formats: in ``MDC maps'' (columns 2 and 4), the color scale is applied independently for {\em each} MDC, while in ``EDC maps'' (columns 3 and 5), it is applied independently for {\em each} EDC\@.  Therefore, an MDC map makes it easy to examine MDC peak positions (red) and MDC peak widths (red-blue distances for a horizontal cut---see panel E for an example) as a function of $\omega$, while an EDC map makes it easy to examine the EDC peak positions (red) and EDC peak widths (red-blue distances for a vertical cut---see panels H, M, or O, for example) as a function of $k$.  Clearly, MDC maps and EDC maps are very informative about MDCs and EDCs, respectively; however, they are, by design, completely irrelevant for EDCs and MDCs, respectively.\\
\hspace*{2em} By following the maximum intensity as a function of $\omega$, we can read off the ``MDC dispersion (MDCD),'' i.e., $\omega$ as a function of the MDC peak position.  Similarly, by tracing the maximum intensity as a function of $k$, we can read off the ``EDC dispersion (EDCD),'' i.e., the EDC peak position as a function of $k$.  Some portions of EDCD and MDCD are marked explicitly in panels D and G.  As for widths, the red-to-blue distance corresponds to the half width at half maximum: marked as $W_L$ and $W_R$ in some panels.  $W_L+W_R$ gives the full width at half maxima.  Note that, generally, $W_L > W_R$ for EDCs, implying asymmetric EDCs, while $W_L \approx W_R$ for MDCs, implying symmetric MDCs.  More acronyms used in this figure: LE = low energy, ME = medium energy, HE = high energy, LEDA = low energy dispersion anomaly, and HEDA = high energy dispersion anomaly, where ``energy'' refers to $-\omega$ in these terms.
}
\label{fig-overview}
\end{figure*} 

We start by discussing the general overview of the data.  Figure \ref{fig-overview} shows the data taken on an optimally doped \bittot{} sample with photon energy 7.5 eV\@.  Here, we clearly visualize the main features of the data: the high energy dispersion anomaly (HEDA; we use ``dispersion anomaly'' in place of the commonly-used ``kink'') and the low energy dispersion anomaly (LEDA).  We also note that MDCs are symmetric while EDCs are asymmetric, a well-known feature from our previous study \cite{matsuyama_phenomenological_2013}.  The detailed quantitative description of the current data, similarly to our previous work \cite{matsuyama_phenomenological_2013}, will be presented in the near future.  For the present discussion, it is sufficient to note that the characteristics of the data taken at 7.5 eV, as noted in this figure, are in good general agreement with the data taken at conventional high energies (25 eV, 33 eV, 55 eV): similar good correspondence between the ``low energy ARPES'' and the ``high energy ARPES'' (see Methods for our definition of these two important terms) have also been noted in previous works \cite{koralek_laser_2006,liu_development_2008,gweon_extremely_2011}.

\begin{figure*}[t] 
\includegraphics[width=0.97\textwidth]{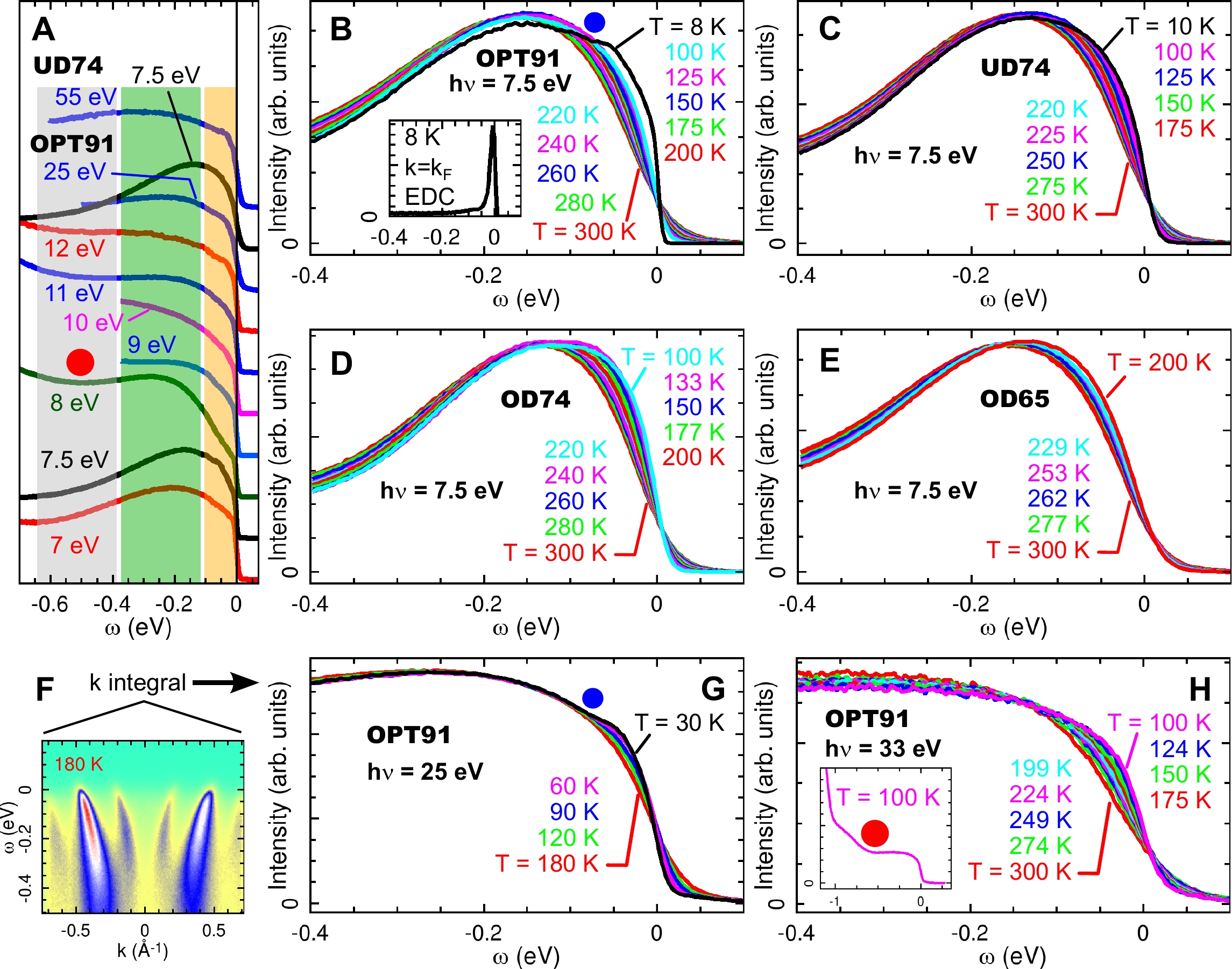}
\caption{
{\bf Nodal many-body density of states.}
The nMBDOS is investigated as a function of photon energy, doping, and temperature.  Panel A shows the photon energy dependence.  All data in panel A are taken at $T \approx 10$ K, except the 7.5 eV data (OPT91) taken at $T \approx 8$ K and the 25 eV data (OPT91) taken at $T = 30$ K\@.  In panels B through E, data taken as a function of temperature are compared for several doping values.  In panels G and H, data taken at two high energy values (25 eV and 33 eV) for OPT91 are reported.  The inset of panel B shows the sharp EDC for this sample, on par with other low energy ARPES data \cite{koralek_laser_2006}.  The temperature in panel G starts from 180 K, which explains why the observed range of the temperature dependence is small, compared to other panels.  In panel F, an ARPES cut that is integrated to give the nMBDOS is shown as an example.  The momentum range of integration varies as a function of photon energy since the total angle range (30 degrees) corresponds to different momentum ranges.  However, the momentum integration range is at least $[-0.15,0.15]$ \AA$^{-1}$ around $k_F$ for all photon energies, suitable for the discussion of nMBDOS in this work.  In fact, the momentum integration range for panels B--E was defined as exactly this range centered around $k_F$, to facilitate the discussion of the electron-hole symmetry, although there is no qualitative change if the integration range is changed slightly.
}
\label{fig-nMBDOS}
\end{figure*} 

In Figure \ref{fig-nMBDOS}, we show the key quantity of interest, the $k$-integrated data.  We shall refer to this quantity as the nMBDOS, as it is proportional to $\int dk A(k, \omega, T)$.

\begin{figure*}[t] 
\includegraphics[width=0.77\textwidth]{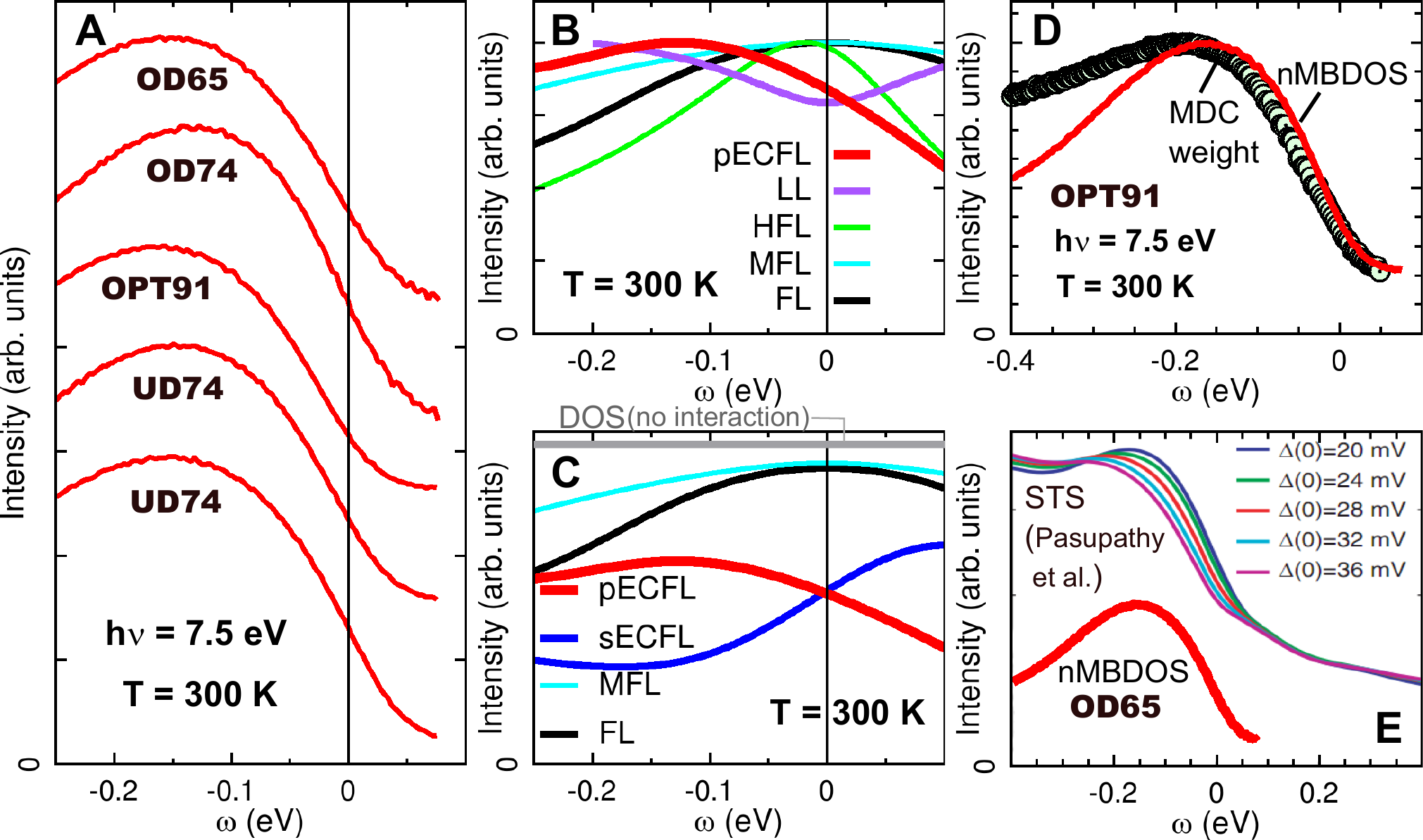}
\caption{
	{\bf Comparison of nMBDOS with theories and STS.} Panel A shows the nMBDOS for $T =$ 300 K, with the energy resolution (6 meV) deconvolved and the Fermi function divided out.  Now the data can be directly compared with $\int dk A(k,\omega,T)$ calculated for different theories.  The top four curves are offset in $y$ for clarity.  The four tick marks on the $y$ axis correspond to the respective zero intensity values for the top four curves.  The top four curves correspond to the data shown in the previous figure.  The bottom curve is the result from another UD74 sample, whose temperature dependent data are not shown here in the interest of space and are in good agreement with the temperature dependence shown in Fig.~\ref{fig-nMBDOS}C.  Shown in panels B and C are theoretical simulations employing two kinds of extremely correlated Fermi liquids (ECFL; pECFL and sECFL) \cite{matsuyama_phenomenological_2013}, Luttinger liquid (LL) \cite{orgad_spectral_2001}, hidden Fermi liquid (HFL) \cite{casey_accurate_2008}, marginal Fermi liquid (MFL) \cite{kaminski_momentum_2005,varma_phenomenology_1989}, and Fermi liquid (FL).  Panel D compares the ``MDC weight'' (see text) and the nMBDOS for the 300 K data for OPT91.  In panel E, we plot our nMBDOS for OD65 (300 K) over the scanning tunneling spectroscopy (STS) data image for an OD62 \bittot{} sample, adopted from Ref.~\onlinecite{pasupathy_electronic_2008}.  For all our data shown in D and E, the energy resolution (6 meV) was deconvolved and the Fermi function was divided out.
}
\label{fig-theories}
\end{figure*} 

We are interested in a few common robust features of the data, found independent of the photon energy used: these are the features that are independent of the ARPES matrix element, and define the intrinsic behavior of the single particle spectral function $A(k, \omega, T)$.  Three such features can be noted in panel A, as marked by three colored rectangular bars.  First, near $\omega = 0$, the nMBDOS  decreases as $\omega$ approaches 0 from the left.  This universal behavior is marked by the pale orange bar.  Second, the nMBDOS shows a maximum at roughly $-\omega \approx 0.2$--$0.3$ eV\@.  This is marked by the green bar.  Third, a shallow minimum (or ``dimple'') is observed around $-\omega \approx 0.5$ eV (red dot; also see the inset of panel H).  By energetics, this feature is associated with the HEDA (see Fig.~\ref{fig-overview}).

Some quantitative differences aside \footnote{These quantitative differences can arise due to several reasons.  One is the ARPES matrix element.  Additionally, the weakening tendency of the features at high photon energy may be related to the extra broadening of peaks for high energy ARPES data \cite{gweon_extremely_2011}.  Also, the tail influence of the intensity from a near-by valence band (the strong turn-on of intensity for $-\omega > 1$ eV in the inset of panel H of Fig.~\ref{fig-nMBDOS}) adds background intensity to weaken the 2nd or 3rd feature.}, these features are robust---this is particularly true for the first feature.

The temperature dependence of the data also confirms this.  A surprisingly large temperature dependence is detected in the data for all doping values examined (panels B--E), regardless of the photon energy used (panels B,G,H).  This dependence is observed to the same degree for both the low energy ARPES data and the high energy ARPES data, pointing to an intrinsic origin.

What could be causing the strong temperature dependent change as $T$ is lowered from 300 K?  A full theory on this is not within the scope of this work; here, we point out key elements that must be present in such a theory.  For this purpose, we focus on the temperature dependence in the low energy regime ($|\omega| \lesssim 0.1$ eV), first.  In this energy scale, the observed temperature dependence can be described as a ``rising up'' of the intensity curve towards $\omega = 0$, as $T$ is lowered, while the curve remains anchored at/near $\omega = 0$.  This trend is far beyond what can be accounted for by the Fermi function alone, as we discuss now.

An important clue can be obtained by observing that, analogous to the HEDA case, a weak dimple appears at the LEDA energy scale (blue dots in panel B and G).  This weak dimple, or small slope change, becomes noticeable at low temperatures, 8 K for panel B and 30 K in panel G.  Concurrently, there is another slope change at about 50 meV, well before the nMBDOS shows a sharp drop at $\omega = 0$, governed by the Fermi function, $f(\omega,T)$.  Having made this observation, one recognizes that our overall temperature dependence can be seen as the gradual softening of these sharp structures as $T$ is raised all the way to 300 K\@.  At the same time, there is weight transfer from low energy to higher energy.  Thus, the mechanism proposed for the low temperature LEDA, whether it involves phonons \cite{zhou_angle-resolved_2007} or magnetic excitations \cite{mook_polarized_1993}, must be an essential ingredient of the theory.  In our previous work \cite{gweon_extremely_2011}, we have pointed out that the Mott-Hubbard physics alone can explain the weak LEDA at high temperatures.  However, facing our new data, we recognize that such an explanation may require temperature-dependent effective parameters, when applied to our full temperature dependent data.  It remains, then, to see whether the strong electron correlation alone can explain the temperature dependence in the normal phase, pointed out in this figure and shown in Fig.~\ref{fig-overview} (panels J,K or M,N) as a change of the strength of the LEDA\@.


We now discuss some characteristics of the data on which we can already shed some theoretical insight.  We shall use the 300 K data for this purpose.  In panel A of Fig.~\ref{fig-theories}, we show the data with the Fermi function divided out \footnote{As noted in the figure caption, the small energy resolution (6 meV) was deconvolved before the dividing out the Fermi function.  However, the deconvolution had little effect on the data since the line shapes were already much broader than 6 meV.} \footnote{We find that this division procedure works well for the low energy ARPES data, but not for the high energy ARPES data, and we attribute this difference to the significant higher order photon contributions for the high energy ARPES data.  For the low energy ARPES, the use of a LiF window to filter out any high order photons resulted in very clean spectra without this issue.}.  In addition to the two key characteristics of the data we noted already, the maximum around $-\omega \approx 0.2$ eV and the decrease of the intensity as $\omega$ approaches zero from below, we can make an important third observation: the intensity keeps decreasing past $\omega = 0$, i.e., the nMBDOS has a {\em negative slope} at $\omega = 0$.

{\em These three characteristics define the ``nMBDOS anomaly,'' the central topic of this work.}  This anomaly is completely separate from the HEDA and the LEDA, as a means to characterize the data, while theoretically these three anomalies may be interconnected.

In panels B and C, we present various theoretical simulations, focusing on OPT91 by choice of parameter values.  The sECFL (simple ECFL) and pECFL (phenomenological ECFL) models are as we defined in our previous work \cite{matsuyama_phenomenological_2013}, where the pECFL model corresponds to the ``MI''-pECFL model of that work.  For the ECFL simulation, the same parameter values of that work are used, except for a small $\eta$ value (50 meV) used here.  The FL theory corresponds to the ``auxiliary'' FL theory used in that work and a previous work \cite{gweon_extremely_2011}.  For the MFL theory, we use $a = 15$ meV (small impurity scattering appropriate for our data) and $b = 0.75$, in the notation of Ref.~\onlinecite{kaminski_momentum_2005}.  The HFL theory is exactly as in Ref.~\onlinecite{casey_accurate_2008}, applied to $T = 300$ K\@.  Lastly, the finite temperature LL theory spectral function is taken from Ref.~\onlinecite{orgad_spectral_2001}, with the anomalous dimension $\alpha = 0.125$ and $v_{c} (k - k_F)$ set equal to $Z_{FL} \varepsilon (k)$ of Ref.~\onlinecite{gweon_extremely_2011} and $v_s / v_c = 1$.  In short, all these theories provide an approximate description for the dispersive peak that crosses the chemical potential at 300 K\@.  In calculations, $A(k,\omega,T)$ was $k$-integrated over the same exact symmetric $k$ range $[-0.15,0.15]$ \AA$^{-1}$ relative to $k_F$, as with the data.

Panel B shows that the general similarity to the data can be found only for the pECFL theory.  The three theories, FL, MFL, and LL, assumed electron-hole symmetry and thus failed to describe the asymmetry of the data upon reflection $\omega \rightarrow -\omega$.  The HFL theory, however, did incorporate electron-hole asymmetry, like pECFL, and it also produced the negative slope at $\omega = 0$.  The disagreement of this theory with the data is, then, only quantitative---its nMBDOS is maximized at about 8 times smaller energy ($-\omega$) than that of the data.

In panel C, we collect results from theories with normalizable spectral functions and compare the absolute magnitude of the nMBDOS with respect to the DOS (for no interaction)\footnote{For the DOS, we use an unlimited $k$-integration window, equivalent to the finite $k$-integration window at low energies, since $A(k, \omega) = \delta (\omega - \varepsilon(k))$.} corresponding to the linear dispersion appropriate for our ARPES cut.  The zero-energy values of MFL and FL were close to the non-interacting value; in fact, they can be shown to converge to the non-interacting value, if the $k$-integration window is enlarged.  On the other hand, the values for ECFL theories are greatly suppressed with respect to the DOS\@.  The discrepancy between the nMBDOS and the DOS at zero energy is well-expected \cite{inoue_systematic_1995} only for a $k$-dependent self energy theory, such as the ECFL or HFL theory.

Of all the models that we examine here, we see that only pECFL and HFL have qualitatively correct behaviors, owing to the two key characteristics: the electron-hole asymmetry and the $k$-dependent self energy.  In terms of energy scale description, pECFL was more accurate.  However, one notes that, in panel B of Fig.~\ref{fig-theories}, HFL does better than pECFL in one regard: the slope at $\omega = 0$ is steeper \footnote{The slope can be made steeper within pECFL if we use about 50 \% smaller $\Delta_0$ value than the previous value \cite{gweon_extremely_2011}, resulting in an excellent agreement with the 7.5 eV data (and HFL).}.

Our findings here go well beyond, and strongly question, the two methods in wide use today: a conventional analysis of MDCs, which relies on the $k$-independence of self energy, and the line shape symmetrization method \cite{norman_phenomenology_1998}, which relies on the electron-hole symmetry.  Regarding the MDC analysis, we already demonstrated that we can interpret MDCs and EDCs on equal footing using the pECFL theory \cite{matsuyama_phenomenological_2013}, and thus our new findings here are on a strong foothold.  In the pECFL analysis of MDCs \cite{matsuyama_phenomenological_2013}, it is the MDC weight that goes with the energy dependence of the nMBDOS, as shown in panel D, and its behavior is traced back to the ``caparison factor'' \cite{shastry_extremely_2011,gweon_extremely_2011}.

Lastly, our finding sheds light on the interpretation of the STS data.  In panel E, we compare the normal state STS data \cite{pasupathy_electronic_2008} for a similarly doped \bittot{} sample as our OD65 sample.  These authors noted a ``hump'' at about $-\omega =$ 0.2 eV, riding on top of an asymmetric background.  As the comparison of our nMBDOS and the hump is quite good, we advance a hypothesis that this hump arises from electrons in the nodal region through the nMBDOS anomaly, while the asymmetric background arises from other regions of the Brillouin zone.  That this hump is found \cite{pasupathy_electronic_2008} to be sensitive to the superconductivity is interesting, in view of the recently proposed importance of the nodal region for the superconductivity \cite{lee_abrupt_2007}.

\vspace{1em}

\begin {acknowledgments} 

{\bf Acknowledgments}
\vspace{0.5em}

We thank B. S. Shastry for stimulating discussions and pointing out Ref.~12.
The SSRL, the ALS, and the BNL are supported by the DOE\@.
This work was partially supported by the faculty research grant at UCSC\@.

\end {acknowledgments} 

\vspace{1em}

{\bf Methods} 
\vspace{0.5em}

Within the standard ``sudden approximation'' theory \cite{hedin__1969}, the ARPES intensity, $I(\kvec,\omega, T)$, as a function of momentum $\kvec$, energy $\omega$ ($\hbar \equiv 1$), and temperature $T$ is given as
\begin{align}
	I(\kvec, \omega, T) &= |M_{if}(\kvec, \vec \epsilon)_\gamma |^2\, A(\kvec, \omega, T)\, f(\omega, T).
\end{align}
Here, $M_{if}$ is the transition dipole matrix element, which involves the initial and final {\em one electron} states of the photoemission process.  Thus, it depends only on $\kvec$ and the properties of the photon (thus the subscript $\gamma$), assuming that one electron states change little as a function of temperature.  $A(\kvec, \omega, T) = \frac{1}{\pi} \Im G(\kvec, \omega, T)$, is the single particle spectral function, where we work with the advanced Green's function, $G$.  $f(\omega, T) = 1/(e^{\omega/(k_BT)} + 1)$ is the Fermi(-Dirac) function.  $A$ and $f$ are independent of the properties of the photon used.  Lastly, all $\kvec$ values used in this work are confined to a one dimensional line (``nodal'' cut; see below), and so we use the symbol $k$, not $\kvec$, for momentum in the main text.


The experimental data were obtained at two synchrotron facilities, the Stanford Synchrotron Radiation Laboratory (SSRL beam line 5-4) and the Advanced Light Source (ALS, beam line 10; 55 eV data).  As a convention, we use the term ``high energy ARPES'' to mean ARPES with photon energy roughly above 10 eV, and ``low energy ARPES'' to mean ARPES with photon energy $\lesssim 10$ eV\@.  At the SSRL, the photon energy is tuned all the way from 7 eV to about 35 eV\@.  The energy resolution (FWHM; Gaussian) at low photon energies was set to 6 meV, while it was set at 15 meV for high photon energies (25 eV, 33 eV, 55 eV)\@.  The angular resolution was 0.3 degrees.  The chemical potential was referenced by measuring a clean gold sample in electrical contact with the sample.  The values of $\omega$ are given relative to the chemical potential, which is defined as zero.  During measurements, the chamber pressure was better than 4$\times 10^{-11}$ Torr.  Temperature dependent measurements were performed by cleaving the crystal at high temperature first, and then cooling the sample down, to prevent sample surface degradation by the heating of the cryostat.  All data are taken along the $(0, 0) \rightarrow (\pi, \pi)$ direction, i.e., the ``nodal direction.''


\bibliographystyle{Science}

\end{document}